\documentclass[12pt]{article}
\usepackage{epsfig}
\usepackage{rotating}

\def\gray{\special{ps: 0.4 setgray}}
\def\black{\special{ps: 0.0 setgray}}

\newcommand{\draft}{
\newcount\timecount
\newcount\hours \newcount\minutes  \newcount\temp \newcount\pmhours
 
\hours = \time
\divide\hours by 60
\temp = \hours
\multiply\temp by 60
\minutes = \time
\advance\minutes by -\temp
\def\hour{\the\hours}
\def\minute{\ifnum\minutes<10 0\the\minutes
            \else\the\minutes\fi}
\def\clock{
\ifnum\hours=0 12:\minute\ AM
\else\ifnum\hours<12 \hour:\minute\ AM
      \else\ifnum\hours=12 12:\minute\ PM
            \else\ifnum\hours>12
                 \pmhours=\hours
                 \advance\pmhours by -12
                 \the\pmhours:\minute\ PM
                 \fi
            \fi
      \fi
\fi
}
\def\fullclock{\hour:\minute}
\begin{centering}
\gray
\begin{sideways}
\font\Hugett  =cmtt12 scaled\magstep2
{\Hugett Draft: \today,\clock}
\end{sideways}
\black
\end{centering}
\vskip -1.7cm
$\phantom{a}$
} 
\catcode`\@=11 
\def\lsim{\mathrel{\mathpalette\@versim<}}
\def\gsim{\mathrel{\mathpalette\@versim>}}
\def\@versim#1#2{\vcenter{\offinterlineskip
        \ialign{$\m@th#1\hfil##\hfil$\crcr#2\crcr\sim\crcr } }}
\catcode`\@=12 

\def\nextline{\hfill\break}

\def\mycomm#1{\nextline\strut\kern-3em{\tt ====> #1}\nextline}

\def\nextline{\hfill\break}
\def\tstrut{\vrule height 1.2em depth 0.5em width 0pt}
\newcommand{\beq}{\begin{equation}}
\newcommand{\eeq}{\end{equation}}
\def\eqref#1{(\ref{#1})}

\begin{document}
\vskip1.5cm
\begin{center}

{\Large\bf The Effect of Doppler Broadening on the $6.3 \ PeV$ $W^-$ Resonance in $\bar{\nu}_e e^-$ Collisions}\\
\tstrut
\vskip 1cm
{\bf Amit Loewy}\footnote{\tt loewy2@post.tau.ac.il}
and
{\bf Shmuel Nussinov}\footnote{\tt nussinov@post.tau.ac.il}
\end{center}
\begin{center}
Raymond and Beverly Sackler School of Physics and Astronomy,\\
Tel Aviv University, Tel Aviv, Israel.\\
\end{center}
\begin{center}
{\bf Sheldon L. Glashow}\footnote{\tt slg@bu.edu}
\end{center}
\begin{center}
Physics Department, Boston University
Boston, MA 02215, USA.
\end{center}
\begin{abstract}
  We calculate the Doppler broadening of the $W^-$ resonance produced
  in $\bar{\nu}_e e^-$ collisions of cosmic anti-neutrinos with
  $E_{\nu}\approx 6.3 \ PeV$ with electrons in atoms up to
  Iron. Revisiting this issue is prompted by recent observations of
  PeV neutrinos by Ice-Cube. Despite its poor energy resolution, the
  $20\%$ Doppler broadening of the resonance due to electronic motions
  can produce observable effects via
  non-linear neutrino absorption near 
the resonance. The attendant suppression of
  the peak cross section allows  $\bar{\nu}_e$ to travel
  correspondingly longer distances. While this effect is unlikely to
  be directly detected in the near future, it may facilitate terrestrial
  tomography at  depths of $\sim 10 \ km$,  complementing
  deeper explorations using the more frequent nuclear interactions at
  lower energies.
\end{abstract}

\newpage
\section{Introduction}
 Some time ago one of us (SLG) noted \cite{Glashow:1960zz} that the $W^-$ boson mediating weak interactions can be resonantly produced in $\bar{\nu}_e e^-$ collisions at $E_R =M_W^2/{2m_e}$ with a peak cross section
\begin{equation}
\sigma (E_R) = 24 \pi \frac{\Gamma_{W \rightarrow e \nu}}{M_W^2 \Gamma_{tot}} \ ,
\end{equation}
where $\Gamma_{W \rightarrow e\nu}$ and $\Gamma_{tot}$ are the ``elastic" and total widths, respectively.
At the time before the $SU(2)_L\times U(1)$ Electroweak standard model was suggested only the combination $\pi\Gamma_W/{M_W^4}=G_F$ was constrained by the Fermi constant. Now we know that $M_W=80 \ GeV$ so that the W Resonance (WR) is at $6.3 \ PeV$, the peak value of the cross section is $5\cdot 10^{-31} \ cm^2$, and the natural width is $\Gamma_W=R_W M_W \sim 0.025M_W$.

The Ultra High Energy (UHE) $\bar{\nu}_e$ required are rare, and the WR has not been seen yet. However, recent observations of  ``PeV" neutrinos in the Ice-Cube neutrino telescope \cite{IC,Aartsen:2013bka,IC2} at the south pole suggest that this may soon happen. Indeed, at its peak the WR cross section is $\sim 500$ times larger than the sum of all neutrino nucleon cross sections \cite{Gandhi:1995tf}. The substantial ratio of the width to the energy of the resonance, $R_W^{lab} =\Gamma^{lab}/E_R=2R_W=0.05$, causes the WR to dominate the region at and around $E_R$  if a significant part of the incident neutrinos are of the $\bar{\nu}_e$ flavor.

\emph{The origin of UHE neutrinos}

UHE cosmic rays produce at the top of the atmosphere pions and kaons that decay into $\mu^-+\bar{\nu}_{\mu}$  followed by $\mu^- \rightarrow{e^-+\bar{\nu}_e+\nu_{\mu}}$, where the anti-neutrino typically carries about ${1/3}$ of the energy of the muon.  However, these anti-neutrinos are unlikely to have energies as high as $6.3\ PeV.$  Pions or muons with Lorentz factors $\gamma \approx {3} E_R/{0.1 GeV}$ have decay lengths of $10^6\ km$ or $10^8\ km$, respectively.  Only a negligible fraction of the pions, and a $100$ times smaller fraction of the muons (which are the only source in this chain of the anti-neutrinos required for the WR) would decay before interacting or hitting the ground.\footnote{Charmed $D^+ D^0; D^- \bar{D}^0$ decay promptly before interacting and about ${10\%}$ of the decays have a final $\nu_e$ or $\nu_{\mu}$. However, the rates of producing very energetic charmed mesons are too small to account for the observed flux.}
  Indeed, the extensive analysis \cite{IC2} of the Ice-Cube collaboration using shower and muon vetoes excluded with high probability atmospheric neutrinos from being the main source of their events.

 Neutrino mixing en-rout to earth generates a $1:1:1$ flavor mix consistent with the measured composition, which contains more ``shower-like" events due to non-$\mu$ neutrinos than ``track events" due to charged current interactions of $\nu_\mu$ and $\bar{\nu}_{\mu}$ producing $\mu^+$ and $\mu^-$. The roughly isotropic angular distribution of the events excludes cosmic rays interacting in the inter-stellar medium of our galaxy from being the main source, so that extra-galactic sources are left as the only likely astrophysical origin of most of the Ice-Cube neutrinos. If the source is indeed extra-galactic, the slowly falling $E^{-2}$ observed spectrum of the UHE neutrinos follows, via Feynman scaling of the $pp\rightarrow \pi^+ + X $ collisions, from the $E^{-2}$ spectrum of the protons which are accelerated a-la Fermi.\footnote{The rate of $p+\gamma\rightarrow{\Delta^+}$ with the subsequent decay of the $\Delta^+$ resonance at $m(\Delta^+)=1240 \ MeV$ into $ n+\pi^+$ depends also on the photon spectrum so the origin of the $ E^{-2}$ spectrum here is less clear.} 

The density of protons and/or photons in the vicinity of the accelerators could be such that the $pp$ and/or the $p\gamma$ collisions can potentially yield the right flux for the Ice-Cube events. Unless the PeV neutrinos originate from the decay of a few PeV dark matter particles in the galactic halo \cite{Esmaili:2013gha,Bella}, or the accelerating ``engine" cuts off before $E_R$, we expect the $E^{-2}$ spectrum to continue beyond $2 \ PeV$ - the highest energy presently observed, and the WR should soon show up.  

The $W$ mass, width and branching ratios agree well with the standard EW model so that discovering the WR will not add to basic particle physics. However, along with nuclear interactions of UHE neutrinos, the WR will help diagnose the source of the UHE neutrinos \cite{Bhattacharya:2011,Bhattacharya:2012fh}, and is also a useful tool for deep earth geological research.

\emph{Interactions in the earth's crust or ice}

It has been suggested \cite{DeRujula} that O$(TeV)$ neutrino beams generated at accelerators could allow large scale earth crust/mantle tomography.  The fact that the earth becomes almost opaque to the UHE neutrinos observed at Ice-Cube, implies that even whole earth tomography may become feasible once many UHE neutrino events with well measured directions and energies are available and the distribution of the incoming flux of neutrinos in the sky is known. Because of the much shorter mean free path (mfp) $l_{mfp}$ of the $6.3 \ PeV$ WR neutrinos:
\begin{equation}
 l_{mfp}(E_R)=(n_e \sigma(E_R))^{-1} \approx 60\  km \ \ \mbox{or} \ \ 20\  km
\end{equation}
in ice or in the earth's crust, they can probe much shorter distances. In the last equation, $n_e$ refers to the average electron number density along the path of the $\bar\nu_e$. Measuring the WR neutrino flux arriving from many directions  will then fix the integrated electron density along these directions.

 A point of key importance is that the linear treatment implicit in the last equation where ``shadowing" effects by higher layers are neglected, is inadequate when the paths traversed by the $\bar{\nu}_e$ are comparable to $l_{mfp}(E_R)$ as the actual flux of WR neutrinos reaching this depth is reduced relative to the original cosmological flux in this direction by absorption. Using optical terminology we have here a ``neutrino absorption line" at $6.3\ PeV$. This line and in particular the effect of Doppler broadening due to the motion of the atomic electrons is the focus of this note.


\section{The Doppler broadening effect} 
\emph{Motivation}

The cross section $\sigma^{(0)}(E)$ for producing the W resonance by an anti-neutrino of energy $E$ impinging on an electron at {\it rest} has the Breit-Wigner (BW) form:
\begin{equation} \label{BWatRest}
\frac{\sigma^{(0)}(E)}{\sigma^{(0)}(E_R)} = \frac{\Gamma_0^2/4}{(E-E_R)^2 + \Gamma_0^2/4}
\end{equation}
where $R_0=\Gamma^{lab}/E_R= 2E_R\Gamma_W/M_W=0.05$ is fixed by the natural $W$ width $\Gamma_W$. The expected rate of interactions $n(E)$  would then be the product $n(E) =\sigma(E)\Phi(E)$ with $\Phi(E)$ the flux of anti-neutrinos at energy $E$. Since the cosmological flux $\Phi^c (E)$ is expected to vary only by a few percents over the resonance, measuring $n(E)$ at various energies around $E_R$ would then yield the expected BW form of $\sigma(E)$. As we will discuss in some detail, the motion of atomic electrons broadens the effective $\sigma(E)$ curve. The effect which increases for heavier elements is moderate - amounting to about $20\%$. 
\newline
Originally \cite {Glashow:1960zz} this effect was estimated by using Bohr's semi-classical model: $\beta= v/c\approx{Z\alpha/n}$, with $Z$ the atomic charge and $n$ the principle quantum number of the electron in question. We improve on this by using the distribution of the velocities calculated by Fourier transforming the various electron wave functions.
As in most neutrino experiments,  we do not know a-priori what is the neutrino energy in each individual event and have to measure it by summing the energies of the particles produced in the event. The large distance between the strings carrying the photo-tubes in Ice-Cube, the large fraction of escaping energy in all leptonic  decays due to the escaping neutrinos especially in $W^- \rightarrow{\mu^-+\bar{\nu}_{\mu}}$ where the escaping muons also carry much energy,  and other possible effects generate a large experimental uncertainty in the measured energy $E$ exceeding the Doppler broadening (DB). Thus, it would seem that the DB is, in the foreseeable future, of academic interest only.

However, neutrino absorption suppresses the flux of neutrinos arriving at the detector after traversing distances comparable to their absorption length $l_{mfp}$. This absorption is particularly relevant for energies near the peak of the BW curve where the mean free path is the shortest. Thanks to the DB we have the peak cross section suppressed by about ${20\%}$, and the relevant penetration depth of the neutrinos will accordingly increase. Note that this increase depends on the specific atomic composition along the neutrino path, which is predominantly that of Oxygen in Ice-Cube experiments but has higher average $Z$ in the earth's crust.  This increase will manifest via $n(E_R,L)$, the {\it total} number of anti-neutrinos with energy $E_R$ which traversed a distance $L$. Note that to find this total number we need not measure the energy of each event. However, to find the distance traveled by the incident neutrino in the ice/crust we need to measure rather precisely directions of final state particles, which by kinematics follow very closely the direction of the initial incoming neutrino. Measuring this direction is most straightforward in   the $O(10\%)$ of the WR events when we have an energetic muon in the final state: when the $W^-$ decays into $\mu^-+\bar{\nu}_{\mu}$ or into $\tau^-+\bar{\nu}_{\tau}$ or $\bar{c}+s$ with the $\tau^-$ or $\bar{c}$ quark decaying into a $\mu^-$.

\emph{Computational details}

We now present some details of the calculation of the DB effect \cite{MScThesis}. Consider a non-relativistic electron with 4-momentum $(m_e+\frac{p^2}{2m_e},{\bf p})$ and an anti-neutrino with 4-momentum $(E,{\bf E})$. The part of the CM energy squared in the WR process that depends on $E$ is
\begin{equation} \label{CMEnergy}
2m_e E + \frac{\bf{p}^2}{m_e} E - 2 \ \bf{p \cdot E} 
\end{equation}
We substitute $|{\bf p}| = m_e \beta$, where $\beta$ is the electron's velocity, and keep only first order terms in ${\bf \beta}$. The scalar product in last term of (\ref{CMEnergy}) contributes a factor of $x=\cos \theta$, where $\theta$ is the angle between the electron and the anti-neutrino velocities. Thus, the BW curve is modified by replacing $E$ with $E(1-\beta x)$ so that:
\begin{equation} 
\sigma(E) = \frac{1}{4\pi} \int d \phi \int d \beta \ F(\beta) \int dx \ \sigma^{(0)} \left[ E(1-\beta x) \right]
\end{equation}
where $F(\beta)$ is velocity distribution of the target electrons. The angular integration over $x$ and $\phi$ yields:
\begin{equation}\label{modBW}
\frac{\sigma(E)}{\sigma^{(0)}(E_R)} = \frac{\Gamma_0}{4E} \int d \beta \ \frac{F(\beta)}{\beta}  \left[ \arctan \frac{2}{\Gamma_0} ( E(1+\beta) - E_R) - \arctan \frac{2}{\Gamma_0} ( E(1-\beta) - E_R) \right]
\end{equation}
 In particular, the ratio of cross sections at $E=E_R$  is
\begin{equation}\label{modBWPeak}
\frac{\sigma(E_R)}{\sigma^{(0)}(E_R)} = \frac{\Gamma_0}{2E_R} \int d \beta \ \frac{F(\beta)}{\beta}  \ \arctan \frac{2E_R\beta}{\Gamma_0} \ .
\end{equation}
To calculate $F(\beta)$ in multi-electron atoms, we Fourier transform the electron's wave function and find the velocity distribution function for the i-th electron:
\begin{equation} \label{SingleEDist}
f_i(\beta) = m_e \int d \Omega_k \ k^2 \ | \Psi_i(k) |^2
\end{equation}
where $\Psi_i(k)$ is the i-th electron wave function in momentum space. Averaging over the $Z$ electrons in the atom we obtain:
\begin{equation}
F(\beta) = \frac{1}{Z} \sum_i f_i(\beta) \ .
\end{equation}
\newline
According to \cite{SlaterBasis} the wave functions can be well approximated by the simple form:
\begin{equation}
\Psi_{n,l}(r,\theta,\phi) = \frac{(2 \xi / a_0)^{n+\frac{1}{2}}}{\sqrt{(2n)!}} r^{n-1} e^{-(\xi r /a_0)} Y_{l,m}(\theta,\phi)
\end{equation}
where $a_0$ is the Bohr radius, and $\xi$ for the (n,l) orbital is 
\begin{equation}
\xi_{n,l} = \frac{Z_{eff}}{n} = \frac{Z- \sigma_{n,l}}{n}
\end{equation}
where $\sigma$ represents the effect of screening of the nuclear charge by the other electrons.
Define $k:=m_e \beta$, and $\mu_{n,l} = \xi_{n,l} /a_0$. Some relevant values of $\mu_{n,l}$ are given in the following table

\begin{center}
\begin{tabular}{|c||c|c|c|c|c|}
\hline
Z & 8 & 12 & 14 & 20 & 26 \\
\hline
\hline
1s & 7.66 & 11.6 & 13.6 & 19.5 & 25.4 \\
2s & 2.25 & 3.7 & 4.5 & 6.9 & 9.3 \\
2p & 2.23 & 3.0 & 5.0 & 8.0 & 11.0 \\
3s &         & 1.1 & 1.6 & 3.2 & 4.6 \\
3p &        &        &  1.4 & 2.9 & 4.3 \\
3d &       &         &        &       &  3.7 \\
4s &       &        &        &  1.1 & 1.4\\
\hline
\end{tabular}
\end{center}
Performing the angular integration part of the Fourier transform we get
\begin{equation}
\Psi_{n,l}({\bf k}) \sim (i)^l Y^*_{l,m}(\Omega_k) \int_0^{\infty} dr r^{n+1} e^{-\mu r} j_l(kr) \ .
\end{equation}
Carrying out the radial integration, and substituting in (\ref{SingleEDist}) we obtain the velocity distributions for electrons in atoms up to $Z=26$.
\begin{eqnarray*}
f_{1s}(k) & = & \frac{32}{\pi} \ \frac{\mu^5 k^2}{(\mu^2 + k^2)^4} \\
f_{2s}(k) & = & \frac{32}{3\pi} \ \frac{\mu^5(3\mu^2k - k^3)}{(\mu^2 + k^2)^6} \\
f_{2p}(k) & = & \frac{512}{3\pi} \ \frac{\mu^7 k^4}{(\mu^2 + k^2)^6} \\
f_{3s}(k) & = & \frac{1024}{5\pi} \ \frac{\mu^7(\mu^3k- \mu k^3)^2}{(\mu^2 + k^2)^8} \\
f_{3p}(k) & = & \frac{1024}{45\pi} \ \frac{\mu^7(5\mu^2 k^2- k^4)^2}{(\mu^2 + k^2)^8} \\
f_{3d}(k) & = & \frac{4096}{5\pi}\  \frac{\mu^9 k^6}{(\mu^2 + k^2)^8} \\
f_{4s}(k) & = & \frac{512}{35\pi}\ \frac{\mu^9(5 \mu^4 k -10 \mu^2 k^3 +  k^5)}{(\mu^2 + k^2)^10} 
\end{eqnarray*}
Using the distribution functions we numerically calculated the integral in (\ref{modBW}). Figure 1 shows the BW curve in $Z=26$. The new curve has a lower peak as expected from (\ref{modBWPeak}). The table below summarizes the effect on the peak of the BW curve (\ref{modBWPeak}) in various elements.
\begin{center}
\begin{tabular}{|c|c|}
\hline
$Z$ & $\sigma(E_R)/\sigma^{(0)}(E_R)$\\
\hline
8 & 0.85 \\
12 & 0.83 \\
14 & 0.78 \\
20 & 0.76 \\
26 & 0.73 \\
\hline
\end{tabular}
\end{center}
\begin{figure}[]
\begin{center}
\includegraphics[height=14cm]{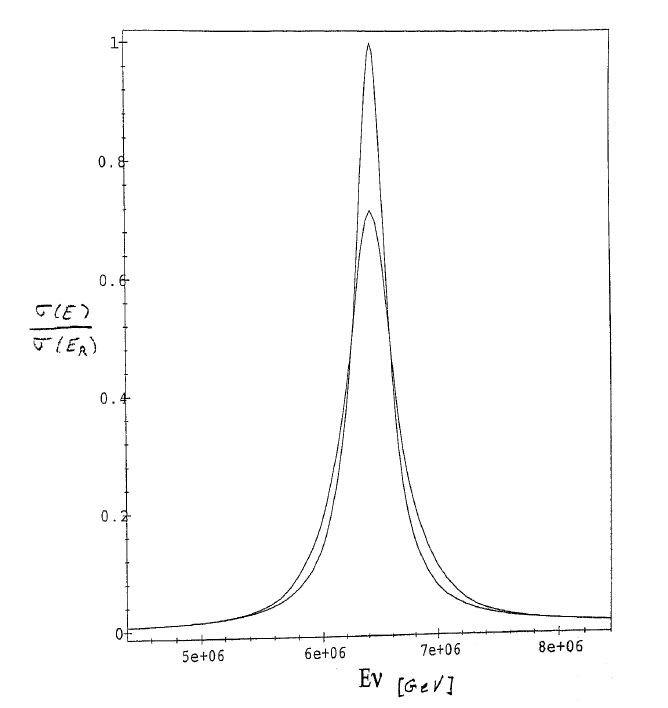}
\end{center}
\caption{Scaled cross sections of the WR on electrons at rest and electrons in $Z=26$. }
\end{figure}
One interesting consequence of the above calculation is that $l_{mfp}$ of the neutrinos does not depend {\it only} on the integrated number of electrons along its path, but thanks to the DB effect, also on the composition of the layers traversed by it. Specifically, going to higher $Z$ materials (which generally also have higher densities) will decrease $l_{mfp}$ more slowly then just in proportion to the electron number density $n_e$.  Unfortunately, it is very hard to test this in Ice-Cube where almost all horizontal long paths with zenith angles approaching $90$ degrees are completely within the ice, and paths with zenith angles slightly larger than $90$ degrees, namely from just under the horizon, are much longer than $60 \ km$, which is the mean free path in the ice.

\section{Summary and conclusions}
 In this short note we have presented a calculation of the Doppler broadening of the $W^-$ resonance produced in collisions of UHE electron anti-neutrinos and atomic electrons. Due to substantial experimental errors in determining the energy of each individual neutrino event, measuring the number of neutrino interactions at any given energy around the resonance is unlikely to reveal the $20\%$ extra broadening - beyond the natural width effect. However, measuring the direction from which the neutrino came can fix the length of path in the ice (or in some future set-ups in rocks as well) which the neutrino had to traverse to reach the interaction point. The ensuing $20\%$ reduction of the peak cross-section and corresponding prolongation of  $l_{mfp}$ may have a measurable impact, and as such may be of interest in geological applications.  
One of the authors (SLG) acknowledges partial support by the Office of
Science of the U.S. Dep't of Energy.


\newpage
\begin{thebibliography}{999}

\bibitem{Glashow:1960zz} 
  S.~L.~Glashow,
  ``Resonant Scattering of Antineutrinos,''
  Phys.\ Rev.\  {\bf 118}, 316 (1960).

\bibitem{IC} IceCube Collaboration, ``Evidence for High-Energy
  Extraterrestrial Neutrinos at the IceCube Detector,'' Science 22
  November 2013: Vol. 342 no. 6161

\bibitem{Aartsen:2013bka} 
  M.~G.~Aartsen {\it et al.}  [IceCube Collaboration],
  ``First observation of PeV-energy neutrinos with IceCube,''
  Phys.\ Rev.\ Lett.\  {\bf 111}, no. 2, 021103 (2013)
  [arXiv:1304.5356 [astro-ph.HE]].

\bibitem{IC2} M.~G.~Aartsen {\it et al.}  [IceCube Collaboration],
 ``Observation of High-Energy Astrophysical Neutrinos in Three Years of IceCube Data,''
  arXiv:1405.5303 [astro-ph.HE].

\bibitem{Gandhi:1995tf} 
  R.~Gandhi, C.~Quigg, M.~H.~Reno and I.~Sarcevic,
  ``Ultrahigh-energy neutrino interactions,''
  Astropart.\ Phys.\  {\bf 5}, 81 (1996)
  [hep-ph/9512364].

\bibitem{Esmaili:2013gha} 
  A.~Esmaili and P.~D.~Serpico,
  ``Are IceCube neutrinos unveiling PeV-scale decaying dark matter?,''
  JCAP {\bf 1311}, 054 (2013)
  [arXiv:1308.1105 [hep-ph]].

\bibitem{Bella} 
 G. Bella, et. al., work in progress.

\bibitem{Bhattacharya:2011} 
  {Bhattacharya}, A. and {Gandhi}, R. and {Rodejohann}, W. and 
	{Watanabe}, A.,
  ``The Glashow resonance at IceCube: signatures, event rates and pp vs. p{$\gamma$} interactions,''
  arXiv:1108.3163v2 [astro-ph.HE].

\bibitem{Bhattacharya:2012fh} 
  A.~Bhattacharya, R.~Gandhi, W.~Rodejohann and A.~Watanabe,
  ``On the interpretation of IceCube cascade events in terms of the Glashow resonance,''
  arXiv:1209.2422 [hep-ph].

\bibitem{DeRujula} 
  A.~De Rujula, S.~L.~Glashow, R.~R.~Wilson and G.~Charpak,
  ``Neutrino Exploration of the Earth,''
  Phys.\ Rept.\  {\bf 99}, 341 (1983).

\bibitem{MScThesis} A.~Loewy, ``Ultra High Energy Neutrino Interactions," M.Sc. Thesis, Tel-Aviv University (1998).

\bibitem{SlaterBasis} E.~Clementi, D.~Raimondi,
The Journal of Chemical Physics {\bf 38} (1963) 2686; {\bf 47} (1967) 1300.






\end{thebibliography}
\end{document}